\documentstyle[12pt,a4]{article}

\begin{document}
  
\def\inbar{\,\vrule height1.5ex width.4pt depth0pt}
\def\IC{\relax\,\hbox{$\inbar\kern-.3em{\rm C}$}}
\def\II{\relax{\rm I\kern-.17em I}}
\def\I1{\relax{\rm 1\kern-.28em l}}
\def\IZ{\relax{\rm /\kern-.18em Z}}
\def\IR{\relax{\rm I\kern-.18em R}}
\font\cmss=cmss12 \font\cmsss=cmss12 at 9pt
\def\Z{\relax\ifmmode\mathchoice {\hbox{\cmss Z\kern-.4em Z}}
{\hbox{\cmss Z\kern-.4em Z}} {\lower.9pt\hbox{\cmsss Z\kern-.4em Z}}
{\lower1.2pt\hbox{\cmsss Z\kern-.4em Z}}\else{\cmss Z\kern-.4emZ}\fi}

\title{\sc Seiberg--Witten Monopole Equations And Riemann Surfaces}
\author{Cihan Sa\c{c}l\i o\~{g}lu$^{1,2}$ \\ 
and \\ 
Serdar Nergiz$^{1}$}
\date{$^{1}$Physics Department, Bo\~{g}azi\c{c}i University \\
80815 Bebek--\.{I}stanbul, Turkey \\
and \\
$^{2}$Physics Department, TUBITAK \\ Marmara Research Center \\
Research Institute for Basic Sciences \\ 41470 Gebze, Turkey}
\maketitle
\vspace*{1cm}
\begin{abstract}
The twice--dimensionally reduced Seiberg--Witten monopole equations 
admit solutions depending on two real parameters $(b,c)$ and an 
arbitrary analytic function $f(z)$ determining a solution of 
Liouville's equation. The $U(1)$ and manifold curvature 2--forms 
$F$ and $R^{1}_{\: 2}$ are invariant under fractional $SL(2,\IR)$ 
transformations of $f(z)$. When $b = 1/2$ and  $c = 0$ and $f(z)$ 
is the Fuchsian function uniformizing an algebraic function whose Riemann 
surface has genus $p \geq 2$ , the solutions, now $SL(2,\IR)$ invariant, 
are the same surfaces accompanied by a $U(1)$ bundle of $c_1 = \pm (p-1)$ and a 
$1$--component constant spinor.
\end{abstract}
\vspace*{3 cm}
\pagebreak
\baselineskip=30pt

\noindent {\bf 1. Introduction: }

The study of the Donaldson invariants of 4--manifolds has become 
considerably simpler in Witten's \cite{art1} new approach 
based on weakly interacting $U(1)$ monopoles. Instead of self--dual 
Yang--Mills 
fields on the manifold, one now deals with a Weyl spinor $\psi$ and a 
$U(1)$ 
connection $A_{\mu}$ , whose interaction is described by the 
Seiberg--Witten monopole equations (SWME)
\begin{equation}\label{1}
\not{\!\! D}_{A} \psi = 0
\end{equation}
\noindent and
\begin{equation}\label{2}
F^{+}_{\mu\nu}=-\frac{i}{4} \psi^{\dag}[\gamma_{\mu},\gamma_{\nu}]\psi~~~,
\end{equation}
\noindent where $F^{+}_{\mu\nu}$ represents the self--dual part of 
$F_{\mu\nu} = \partial_{\mu} A_{\nu} - \partial_{\nu} A_{\mu}$ .

An immediate consequence of (\ref{1}) and (\ref{2}), obtained by using the 
Weitzenbock formula, is that nontrivial and nonsingular 
solutions are possible only when the scalar curvature $R$ is negative in 
some regions of the manifold. Witten's vanishing theorem \cite{art1} , 
based on an integrated version of the same argument, establishes that 
$R \geq 0$ solutions are not only singular but also non-$L^2$. 
Nevertheless, such solutions may be worth studying if they are related to 
field configurations of physical interest. For example, Freund 
\cite{art2} has exhibited a once--dimensionally reduced solution in 
$\IR^{4}$, also found earlier by G\"{u}rsey \cite{art3} in another 
setting, corresponding to a singular Dirac monopole. Still working in 
$\IR^{4}$, we have shown \cite{art4} that the SWME allow singular 
multi-vortices and singular $\varphi^4$--kink solutions (the standard 
non--singular versions are obtained by passing from $(++++)$ to 
$(++--)$ signature) when dimensionally reduced to $d = 2$ and $d = 1$, 
respectively.

In this paper we generalize our search for $d = 2$ solutions by 
introducing a non--flat metric in two dimensions. The curvature can now 
assume negative values, and non--singular solutions become possible. As 
will be seen below, a subset of these solutions are nothing but Riemann 
surfaces of genus $p > 1$ with Chern number $p-1$ ! These topological 
features are found to be invariant under a wide class of transformations 
on the metric and other fields. This is perhaps a reflection of the 
connection, pointed out by Olsen \cite{art5}, between twice--dimensionally 
reduced SWME and certain $2d$ topological field theories \cite{art6}.

Let us briefly summarize the paper. In section 2, we write down the SWME 
for a product manifold ${\cal M}_4 = {\cal M}_2 \otimes S^1 \otimes S^1$, 
where the radii of the circles have been taken to zero (of course, as long as we 
keep only the lowest modes, we could also consider 
${\cal M}_2 \otimes S^2$). We then specialize to a simple Ansatz which 
reduces the SWME to a covariant form of the Poisson--Boltzmann equation 
appearing in non--singular multivortex solutions given in  
\cite{art7} and \cite{art8}. In section 3, we present a special set 
of solutions which depend on two parameters $b$ and $c$ and an 
arbitrary analytic function $g(z)$ characterizing a solution of 
Liouville's equation. In section 4, we show that $b$ determines the sign 
of the curvature $R^{1}_{\: 212}$ of ${\cal M}_2$ (and thus also its 
topology, through the Gauss--Bonnet formula), while changing $c$ only 
effects $U(1)$ gauge transformations on $\psi$ and $A_{\mu}$,  
simultaneously  with a conformal transformation on the metric 
$h_{\alpha \beta}$ of 
${\cal M}_2$. In section 5, the same transformation property is seen to 
hold when $g(z)$ is subjected to the fractional version of $SU(1,1)$ 
transformations
$\tilde{g} = (\alpha g + \beta)/(\overline{\beta} g + \overline{\alpha})$  
with $\alpha \overline{\alpha} - \beta \overline{\beta} = 1$. 
In section 6, we study in greater detail the 
solutions that correspond to specific values of $b$ and $c$; the case 
$b = 1$ is seen to revert to our earlier \cite{art4} singular 
multivortices in $\IR^4$ , while $b > 1$ duly leads to singular 
$R \geq 0$ solutions. On the other hand, $b = 1/2$ , $c = 0$ yields a 
constant negative curvature ${\cal M}_2$ ; according to a well-known 
theorem, one can obtain all genus $p \geq 2$ Riemann surfaces by dividing 
this ${\cal M}_2$ by appropriate discrete groups. This `division' or 
tesellation of the hyperboloid by $4p$-gons with geodesic edges 
corresponds to expressing $g(z)$ in terms the Fuchsian functions 
introduced by Poincar\'{e}. Remarkably, a recent analysis \cite{art9} of 
the Seiberg--Witten Ansatz \cite{art10} in its original physics context 
(quark confinement via a monopole condensate) reveals that the essence of 
the approach consists in identifying the function ${\cal F}''(a)$ 
governing the effective supersymmetric Lagrangian with the simplest 
Fuchsian function, which happens to be the inverse of the elliptic modular 
function! With the choice of a suitable Fuchsian function for $g(z)$, 
${\cal M}_2$ becomes a Riemann surface of Euler characteristic 
$\chi = 2 - 2p$ accompanied by a $U(1)$ bundle of Chern number 
(physically, the number of vortices) $c_1 = p-1$ and a constant Weyl spinor. 
Section 7 ends the paper with concluding remarks concerning Riemann 
surfaces, Integrable Systems and the Seiberg--Witten equations . 

\pagebreak

\noindent {\bf 2. Reduction of the SWME to d = 2 :}

Any ${\cal M}_2$ metric $h_{\alpha \beta}$ can be written in the 
conformally 
flat form $h_{\alpha \beta} = e^{2\phi} \delta_{\alpha \beta}$. After 
dimensional reduction, the basis $1$--forms may be taken as 
\begin{equation}\label{3}
e^1 = e^{\phi}dx^1~~,~~e^2 = e^{\phi}dx^2~~,~~e^3 = dx^3~~,~~e^4 = dx^4~~,
\end{equation}
\noindent where $\phi = \phi(x^1,x^2)$. The Cartan structure equations 
yield
\begin{equation}\label{4}
\omega^{1}_{\: 2} = (\partial_2 \phi)dx^1 - (\partial_1 \phi)dx^2~~.
\end{equation}
\noindent The curved-index $\gamma$--matrices are
\begin{equation}\label{5}
\gamma^{\mu} = \gamma^{a} E^{\: \mu}_{a}~~.
\end{equation}
\noindent Here $E_{a}^{\mu} e_{\mu}^{b} = \delta_{a}^{b}$ ; thus
\begin{equation}\label{6}
E_1 = e^{-\phi}\partial_1~~,~~E_2 = e^{-\phi}\partial_2~~,
~~E_3 = \partial_3~~,~~E_4 = \partial_4~~.
\end{equation}
\noindent We take the flat--space $\gamma$--matrices as
\begin{eqnarray}\label{7}
\gamma^1 &=& \tau_1 \otimes \sigma_1 = 
             \left( \begin{array}{cc} {\bf 0} & {\bf \sigma}_1 \\
             {\bf \sigma}_1 & {\bf 0} \end{array}\right)~~,~~
           \gamma^2 = \tau_1 \otimes \sigma_2 = 
           \left( \begin{array}{cc} {\bf 0} & {\bf \sigma}_2 \\
           {\bf \sigma}_2 & {\bf 0} \end{array}\right)~~, \nonumber \\
\gamma^3 &=& \tau_1 \otimes \sigma_3 =
             \left( \begin{array}{cc} {\bf 0} & {\bf \sigma}_3 \\
             {\bf \sigma}_3  & {\bf 0} \end{array}\right)~~,~~
           \gamma^4 = \tau_2 \otimes \I1 = 
           \left( \begin{array}{cc} {\bf 0} & -i\I1 \\
           i\I1 & {\bf 0} \end{array}\right)~~, \nonumber \\
\gamma^5 &\equiv& 
         \gamma^4 \gamma^1 \gamma^2 \gamma^3 = \left( \begin{array}{cc}
              \I1 & {\bf 0} \\
              {\bf 0} & -\I1 \end{array}\right)~~.
\end{eqnarray}
\noindent These act on the Weyl spinor 
$\psi^T = (\psi_1 (x^1 , x^2),\psi_2 (x^1 , x^2),0,0)$. One of the SWME is 
the Dirac equation
\begin{equation}\label{8}
\not{\!\! D}_{A} \psi = 0~~,
\end{equation}
\noindent where
\begin{equation}\label{9}
\not{\!\! D}_{A} = \gamma^a E^{\mu}_{a}
(\partial_{\mu}+iA_{\mu}+\frac{1}{8}
\omega^{bc}_{\mu}[\gamma_b,\gamma_c])~~.
\end{equation}
\noindent Written explicitly,  (\ref{9}) becomes the pair of equations
$$e^{\phi} (iA_3 - A_4) \psi_1 + 
(\partial_1 - i\partial_2 + iA_1 + A_2 + \frac{1}{2}\partial_1 \phi -
 \frac{i}{2}\partial_2 \phi)\psi_2 = 0 
\eqno(10.a)$$
\noindent and
$$e^{\phi} (-iA_3 - A_4) \psi_2 + 
(\partial_1 + i\partial_2 + iA_1 - A_2 + \frac{1}{2}\partial_1 \phi +
 \frac{i}{2}\partial_2 \phi)\psi_1 = 0~~.~~
\eqno(10.b)$$
\setcounter{equation}{10}
\noindent It is of course understood that $A_{\mu} = A_{\mu}(x^1 , x^2)$ 
only. The remaining SWME are
\begin{equation}\label{11}
F^{+}_{\mu\nu}=-\frac{i}{4} \psi^{\dag} [\gamma_{\mu} , \gamma_{\nu}]
\psi~~~,
\end{equation}
\noindent where 
\begin{equation}\label{12}
F^{+}_{\mu\nu} = \frac{1}{2}(F_{\mu\nu} + 
\frac{1}{2}\epsilon_{\mu \nu \alpha \beta}F^{\mu\nu})~~,
\end{equation}
\noindent and
\begin{equation}\label{13}
F_{\mu\nu} = \partial_{\mu} A_{\nu} - \partial_{\nu} A_{\mu}
\end{equation}
\noindent as usual. The three components in (\ref{11}) yield
$$\partial_1 A_2 - \partial_2 A_1 = e^{2\phi}
(|\psi_1|^2 - |\psi_2|^2)~~,~~~
\eqno(14.a)$$
$$\partial_2 A_3 + e^{-2\phi}\partial_1 A_4 = e^{\phi}
(\overline{\psi_1}\psi_2 + \overline{\psi_2}\psi_1 )~~,~
\eqno(14.b)$$
\noindent and
$$\partial_2 A_4 - e^{-2\phi}\partial_1 A_3 = ie^{\phi}
(\overline{\psi_1}\psi_2 - \overline{\psi_2}\psi_1 )~~.
\eqno(14.c)$$
\setcounter{equation}{14}

The most general form of the SWME depending on two variables is 
represented by the set (10) and (14), involving $5$ real 
$(A_{\mu} , \phi)$ and two complex $(\psi_1 , \psi_2)$ functions. In order 
to obtain a solution, we will consider progressively more restricted 
Ans\"{a}tze. We start by looking for solutions with $A_3 = A_4 = 0$ , 
which, by (14.b) and (14.c), implies that either $\psi_1 = 0$ or 
$\psi_2 = 0$. Let us begin with the case $\psi_2 = 0$. The SWME are now 
reduced to 
\begin{equation}\label{15}
(\partial_1 + i\partial_2 + iA_1 - A_2 + \frac{1}{2}\partial_1 \phi +
 \frac{i}{2}\partial_2 \phi)\psi_1 = 0
\end{equation}
\noindent and
\begin{equation}\label{16}
\partial_1 A_2 - \partial_2 A_1 = e^{2\phi}|\psi_1|^2 ~~.
\end{equation}
\noindent It is useful to write
\begin{equation}\label{17}
\psi_1 = a\exp{(\omega_x +i \omega_y)}~~,
\end{equation}
\noindent where $\omega_x$ and $\omega_y$ are real functions and $a$ is a 
positive constant with the dimensions of inverse length. Note that because 
the vacuum expectation value of the Higgs field from the original twisted 
supersymmetric Yang--Mills theory has \cite{art11} been incorporated into 
the equations, the spinor field dimension is no longer the usual 
$[L]^{-3/2}$. Dividing (\ref{15}) by $\psi_1$ , applying 
$\partial_1 - i\partial_2$ on the result, separating real and imaginary 
parts and using (\ref{16}) lead to the pair of equations 
\begin{equation}\label{18}
(\partial_1 \partial_1 +  \partial_2 \partial_2)
(\omega_x + \phi /2) = a^2\exp{(2\omega_x + 2\phi)}
\end{equation}
\noindent and 
\begin{equation}\label{19}
(\partial_1 \partial_1 +  \partial_2 \partial_2)\omega_y = 
-(\partial_1 A_1 + \partial_2 A_2)~~.
\end{equation}
\noindent Making use of 
\begin{equation}\label{20}
R^{1}_{\: 2} = d\omega^{1}_{\: 2} = - (\partial_1 \partial_1 \phi + 
\partial_2 \partial_2 \phi) e^{-2\phi} e^1 \wedge e^2
\end{equation}
\noindent and 
\begin{equation}\label{21}
\sqrt{\det (h_{\alpha \beta})} = e^{2\phi}
\end{equation}
\noindent  (\ref{18}) can be written in the form
\begin{equation}\label{22}
(\omega_x)^{;\mu}_{\: ;\mu} - \frac{1}{2}R = a^2 e^{2\omega_x}~~,
\end{equation}
\noindent  where $R$ is the scalar curvature.

(\ref{22}) represents a covariant generalization of some well--known 
non--linear field equations in two dimensions. For instance, when the 
metric is flat $(\phi = 0)$, (\ref{22}) reduces to the Liouville equation, 
whose solutions are singular along a closed curve in the $x^1x^2$--plane; 
this is the case already examined in \cite{art4}. For constant negative 
curvature, (\ref{22}) becomes the covariant form of the Poisson--Boltzmann 
equation which arises in a number of seemingly unrelated contexts, e.g., 
in the Debye-H\"{u}ckel theory \cite{art8} of electrolytic solutions, as 
well as in the parametrization of non-interacting two-dimensional 
multivortex solutions without singularities \cite{art7}. Since the 
constant term in the Poisson--Bolzmann equation helps cure the Liouville 
singularities in the multivortex problem, one may also expect 
(\ref{22}) to yield non--singular solutions for appropriate negative 
curvature metrics $h_{\alpha \beta}$ .

\noindent {\bf 3. A 2--parameter class of special solutions: }

Obtaining the general solution of (\ref{22}) for {\em a given} 
metric $h_{\alpha \beta}$ is a highly non-trivial task. However, a special 
class of solutions is readily available. Let us first introduce a 
parameter $b$ and write 
\begin{equation}\label{23}
\omega_x + \phi /2 = b (\omega_x + \phi) + 
[(1-b)\omega_x + (1-2b)\phi /2]
\end{equation}
\noindent and then impose
\begin{equation}\label{24}
(\partial_1 \partial_1 +  \partial_2 \partial_2)
[(1-b)\omega_x + (1-2b)\phi /2] = 0~~.
\end{equation}
\noindent Next, define the dimensionless complex coordinates 
\begin{equation}\label{25}
z \equiv \frac{a}{\sqrt{b}}(x^1 + ix^2)~~,~~
\overline{z} \equiv \frac{a}{\sqrt{b}}(x^1 - ix^2)~~,
\end{equation}
\noindent in terms of which (\ref{18}) (which is another form of 
(\ref{22})) becomes 
\begin{equation}\label{26}
4\partial_{z} \partial_{\overline{z}}(\omega_x + \phi) = 
e^{2(\omega_x + \phi)}
\end{equation}
\noindent since (\ref{24}) holds. Thus we end up with the Liouville 
equation for the combination $(\omega_x + \phi)$, in contrast to the 
Liouville equation satisfied by $\omega_x$ alone in the $\IR^{4}$ case. 
The solution of (\ref{26}) is due to Liouville \cite{art12}; it is 
given by
\begin{equation}\label{27}
\omega_x + \phi=\frac{1}{2}\ln(4g'\overline{g}')-\ln(1-g\overline{g})~~,
\end{equation}
\noindent where $g(z)$ denotes an arbitrary analytic function, 
$\overline{g}$ 
its antianalytic complex conjugate and $g'= dg/dz$. One 
satisfies (\ref{24}) automatically by taking 
\begin{equation}\label{28}
(1-b)\omega_x + (1-2b)\phi /2 = c\ln(4g'\overline{g}')^{1/2}
\end{equation}
\noindent as $\partial_{z} \partial_{\overline{z}}$ annihilates the 
right--hand side of (\ref{28}); $c$ is an additional arbitrary parameter. 
Now solving for $\phi$ and $\omega_x$ from (\ref{28}) and (\ref{27}), we 
find 
\begin{equation}\label{29}
\omega_x = [2(b+c)-1] \ln\sqrt{(4g'\overline{g}')}+
(1-2b)\ln(1-g\overline{g})
\end{equation}
\noindent and 
\begin{equation}\label{30}
\phi = 2[1-(b+c)] \ln\sqrt{(4g'\overline{g}')}-
2(1-b)\ln(1-g\overline{g})~~.
\end{equation}
\noindent From $|\psi_1| = a\exp{\omega_x}$, we have
\begin{equation}\label{31}
|\psi_1| = 
a\frac{(4g'\overline{g}')^{b+c-1/2}}{(1-g\overline{g})^{2b-1}}~~.
\end{equation}
\noindent It is then natural to take
\begin{equation}\label{32}
\psi_1 = 
ae^{\omega_x + i\omega_y} = 
a\frac{(2g')^{2b+2c-1}}{(1-g\overline{g})^{2b-1}}~~.
\end{equation}
\noindent which  means we have chosen
\begin{equation}\label{33}
\omega_y = i(b+c-1/2)\ln(\overline{g}'/g')~~.
\end{equation}
Solving for the $U(1)$ connection from (\ref{15}), we obtain
$$
A_1 = \frac{ia}{2\sqrt{b}}
\left\{
[1-(b+c)] 
\left(
\frac{\overline{g}''}{\overline{g}'} - \frac{g''}{g'}
\right) 
+ 2b\frac{(g\overline{g}'- g'\overline{g})}{(1-g\overline{g})}
\right\}
\eqno(34.a)
$$
\noindent and
$$A_2 = \frac{a}{2\sqrt{b}}
\left\{ 
[1-(b+c)] 
\left( \frac{\overline{g}''}{\overline{g}'} + \frac{g''}{g'} 
\right) 
+ 2b \frac{(g\overline{g}'+ g'\overline{g})}{(1-g\overline{g})}
\right\}~~.
\eqno(34.b)$$
\setcounter{equation}{34}
\noindent With these $(A_1 , A_2)$ and $\omega_y$ , equation (\ref{19}) is 
seen to hold with both sides vanishing. One may now go back and check that 
(\ref{30}), (\ref{32}) and (34) solve the original SWME.

It is useful to express the $U(1)$ connection as a $1$-form. This gives 
\begin{equation}\label{35}
A = \frac{i}{2}[1-(b+c)]d\ln(\overline{g}'/g')+ 
ib\frac{(gd\overline{g}-\overline{g}dg)}{(1-g\overline{g})}~~.
\end{equation}
\noindent The spin--connection $1$--form is 
\begin{equation}\label{36}
\omega^{1}_{\: 2}= i[1-(b+c)]d\ln(g'/\overline{g}')+ 
2i(1-b)\frac{(\overline{g}dg - gd\overline{g})}{(1-g\overline{g})}~~.
\end{equation}
One may define complex basis $1$--forms via
\begin{equation}\label{37}
e^z \equiv e^1 + ie^2 = \frac{\sqrt{b}}{a}dz
\frac{(4g'\overline{g}')^{1-(b+c)}}{(1-g\overline{g})^{2-2b}}
\end{equation}
\noindent and
\begin{equation}\label{38}
e^{\overline{z}} \equiv e^1 - ie^2 = \frac{\sqrt{b}}{a}d\overline{z}
\frac{(4g'\overline{g}')^{1-(b+c)}}{(1-g\overline{g})^{2-2b}}
\end{equation}
\noindent The curvature $2$--form for the $U(1)$ connection
\begin{equation}\label{39}
F = dA =  2ib\frac{ d g \wedge d \overline{g}}{(1-g\overline{g})^2}
\end{equation}
\noindent and the manifold 
\begin{equation}\label{40}
R^{1}_{\: 2} = d\omega^{1}_{\: 2} =  
-4i(1-b)\frac{dg \wedge d \overline{g}}{(1-g\overline{g})^2}
\end{equation}
\noindent are seen to be proportional to each other. Stripping away the 
basis $1$--forms, we have
\begin{equation}\label{41}
R^{1}_{\: 212} = -2a^2\left( \frac{1-b}{b} \right) 
\frac{(4g'\overline{g}')^{2(b+c)-1}}{(1-g\overline{g})^{4b-2}}
\end{equation}
\noindent and 
\begin{equation}\label{42}
F_{12} = 
\frac{4a^2g'\overline{g}'}{(1-g\overline{g})^2}
\end{equation}
\noindent for the Riemann and $U(1)$ field strength tensors, respectively. 
In two dimensions $R^{1}_{\: 212}$ of course coincides with the scalar 
curvature $R$. The equation (\ref{41}) shows that 
\\
$$\begin{array}{llll}
(i)~~~~~~& R<0~~~~~& {\rm for}~~~ & 1>b>0~~,\\
(ii)~~~~~& R>0~~~~~& {\rm for}~~~ & b>1~~,\\
(iii)~~~~& R=0~~~~~& {\rm for}~~~ & b=1~~.\\
\end{array}\eqno(43)$$ 
\setcounter{equation}{43}
\\
\noindent One might think (\ref{41}) also allows $R>0$ when $b<0$, but a 
glance at (\ref{25}) reveals that negative values of $b$ lead to $z$ and 
$\overline{z}$ being multiplied by $\sqrt{-1}$, resulting in a signature 
change from $(++++)$ to $(--++)$.  This would also require changes in the 
Diracology; thus our framework restricts us to $b>0$ as (\ref{25}) shows 
$b=0$ is also to be excluded.

The three possibilities in (43) may remind the reader of a fundamental 
theorem in two dimensional geometry and topology which states that all 
Riemann surfaces can be derived from the sphere $S^2$, the complex 
plane $\IC$ or the constant negative curvature hyperboloid $H^2$; in 
particular, the genus $p=1$ torus and $p\geq 2$ surfaces are obtained by 
dividing $\IC$ and $H^2$ by appropriate discrete groups. Conformal 
transformations can be used to generate surfaces homeomorphic to the 
above.  We will indeed see later that the genus $p>1$ surfaces emerge 
as a specific subset of our solutions for particular choices of the analytic 
function $g(z)$ and the parameters $b$ and $c$.

\pagebreak

\noindent {\bf 4. The significance of the parameters $\boldmath b$ and 
$\boldmath c$ : }

Using the Gauss--Bonnet theorem expression
\begin{equation}\label{44}
\frac{1}{2\pi}\int R^{1}_{\: 2} = \chi
\end{equation}
\noindent for the Euler characteristic $\chi$ and the definition of the 
first Chern number
\begin{equation}\label{45}
\frac{1}{2\pi}\int F = c_1
\end{equation}
\noindent together with (\ref{39}) and (\ref{40}), we see that the 
topological invariants depend on $b$ but not on $c$. One therefore expects 
that the transformation 
\begin{equation}\label{46}
c \rightarrow c + \delta c
\end{equation}
\noindent can at most result in a combination of a $U(1)$ gauge 
transformation and a conformal change of the metric $h_{\alpha \beta}$ . 
Applying (\ref{46}) on (\ref{32}) , (\ref{35}) , (\ref{36}) and (\ref{41}) 
one indeed finds
\begin{equation}\label{47}
A \rightarrow A - \frac{i}{2} \delta c \; d\ln(\overline{g}'/g')~~,
\end{equation}
\begin{equation}\label{48}
\omega^{1}_{\: 2} \rightarrow \omega^{1}_{\: 2} + i \delta c \;
d\ln(\overline{g}'/g')~~,
\end{equation}
\begin{equation}\label{49}
R^{1}_{\: 212} \rightarrow (4\overline{g}'g')^{2\delta c} \:
R^{1}_{\: 212}~~,
\end{equation}
\begin{equation}\label{50}
\det (h_{\alpha \beta})^{1/2} \equiv h^{1/2} = e^{2\phi} \rightarrow 
(4\overline{g}'g')^{-2\delta c} \: e^{2\phi}~~,
\end{equation}
\noindent and
\begin{equation}\label{51}
\psi_1 \rightarrow (2g')^{2\delta c} \: \psi_1~~.
\end{equation}
\noindent At first sight it might appear that (\ref{51}) is not the 
anticipated $U(1)$ transformation for a spinor wavefunction until one 
realizes, on both physical and mathematical grounds, that in the presence 
of a non--flat ${\cal M}_2$ metric $h_{\alpha \beta}$ , the usual $U(1)$ 
transformation must apply to the {\em spinor density} 
$\psi_1 h^{1/4}$ , which does transform in the expected way
\begin{equation}\label{52}
\psi_1 h^{1/4} \rightarrow \left( g'/\overline{g}' 
\right)^{\delta c} \: \psi_1 h^{1/4}~~.
\end{equation}

\noindent {\bf 5. $\boldmath SU(1,1)$ symmetry of the solutions : }

The reader may have noticed that there is a bigger symmetry group, beyond 
the $1$--parameter group noted in section 4, that leaves the $2$--form 
$dg \wedge d\overline{g} / (1-g\overline{g})^2$ (and hence the topology ) 
invariant. This consists of the well--known fractional transformations 
\begin{equation}\label{53}
\tilde{g} = \frac{\alpha g+\beta}{\overline{\beta}g+\overline{\alpha}} 
~~{\rm with}~~|\alpha|^2 - |\beta|^2 = 1~~.
\end{equation}
\noindent Thus 
$\left( 
\begin{array}{cc} \alpha           & \beta \\
                  \overline{\beta} & \overline{\alpha} \end{array} 
\right) \in SU(1,1)$ 
and $|g| \leq 1$  is mapped into $|\tilde{g}| \leq 1$ . The effect of 
(\ref{53}) on the fields is given by 
\begin{equation}\label{54}
A \rightarrow A - i(2b+c-1) \, d\ln 
\left(
\frac{\overline{\beta}g + \overline{\alpha}}{\beta \overline{g} + \alpha} 
\right)~~,
\end{equation}
\begin{equation}\label{55}
\omega^{1}_{\: 2} \rightarrow \omega^{1}_{\: 2} + 2i c \,  d\ln
\left(
\frac{\overline{\beta}g + \overline{\alpha}}{\beta \overline{g} + \alpha} 
\right)~~,
\end{equation}
\begin{equation}\label{56}
\psi_1 \rightarrow \psi_1  \, 
\frac{|\overline{\beta}g +\overline{\alpha}|^{(4b-2)}}{(\overline{\beta}g 
+ \overline{\alpha})^{(4b+4c-2)}}~~,
\end{equation}
\begin{equation}\label{57}
\psi_1 h^{1/4} \rightarrow \psi_1 h^{1/4} \, 
\left\{
\frac{\beta \overline{g} + \alpha}{\overline{\beta}g + \overline{\alpha}}
\right\}^{2b+2c-1}~~,
\end{equation}
\begin{equation}\label{58}
h^{1/2} = e^{2\phi} \rightarrow 
|\overline{\beta}g + \overline{\alpha}|^{8c} \, e^{2\phi}~~,
\end{equation}
\noindent and
\begin{equation}\label{59}
R^{1}_{\: 212} \rightarrow |\overline{\beta}g +\overline{\alpha}|^{-8c} \,
R^{1}_{\: 212}~~,
\end{equation}
\noindent We note again the simultaneous occurence of $U(1)$ gauge 
transformations on $A$ and $\psi$, together with conformal transformations 
on $h_{\alpha \beta}$ , $\omega^{1}_{\: 2}$ and $R^{1}_{\: 212}$ . 
(\ref{47})--(\ref{52}) and (\ref{54})--(\ref{59}) are reminiscent of 
Hermann Weyl's early but mistaken attempt to obtain the form of the 
electromagnetic coupling from a requirement of local scale invariance; 
here Weyl's original `Eichinvarianz' and the local $U(1)$ symmetry are 
actually linked! 

The above fractional $SU(1,1)$ transformations on $g(z)$ are isomorphic 
to $SL(2,\IR)$ fractional transformations on the analytic function 
$f(z)$ defined by 
\begin{equation}\label{60}
g(z)=\frac{f(z)-i}{f(z)+i}~~.
\end{equation}
\noindent The relation (\ref{60}) and its inverse map the region 
$|g| \leq 1$ to the upper half $f$--plane ${\rm Im}\,f \geq 0$ and vice versa.
The counterpart of (\ref{53}) is now 
\begin{equation}\label{61}
\tilde{f} = \frac{Af+B}{Cf+D}~~,
\end{equation}
\noindent where, with
\begin{equation}\label{62}
\alpha \equiv p+iq~,~~\beta \equiv r+is~,~~p,q,r,s \in {\boldmath R}~~,
\end{equation}
\noindent one has
\begin{equation}\label{63}
p = \frac{1}{2}(A+D)~,~~q = \frac{1}{2}(B-C)~,~~r = \frac{1}{2}(A-D)~,~
s = \frac{-1}{2}(B-C)
\end{equation}
\noindent and
\begin{equation}\label{64}
|\alpha|^2 -|\beta|^2 = AD - BC = 1~~.
\end{equation}
\noindent The function $f$ is more convenient to work with than the 
function $g$ when 
one wishes to focus on the modular subgroup $SL(2,\Z)$ of 
$SL(2,\IR)$.

\noindent {\bf 6. Special values of $\boldmath b$ and $\boldmath c$ and 
Riemann surfaces : }

Having seen that $c$ has no topological significance , we can set $c = 0$ 
and concentrate on $b$, which must be greater than zero (recall the 
discussion at the end of section 3 and equation (\ref{25})). It is easy to 
see that $b$ can only assume a restricted set of values when $c = 0$. Let 
us put $g = |g| e^{i \gamma}$ and take $\alpha = 0$, $\beta = 1$. Although 
now $|\alpha|^2 - |\beta|^2  = -1$, this is still an admissible $U(1)$ 
gauge transformation under which (\ref{57}) becomes 
\begin{equation}\label{65}
\psi_1 h^{1/4} \rightarrow \psi_1 h^{1/4} e^{-2i\gamma (2b -1)}~~.
\end{equation}
\noindent Demanding single valuedness of $\psi_1 h^{1/4} $ under 
$\gamma \rightarrow \gamma + 2\pi $, we get
\begin{equation}\label{66}
b = \frac{n}{4} + \frac{1}{2}~~,~~ n \in \Z~~,
\end{equation}
\noindent the same conclusion also follows from (\ref{32}). 

We thus have the three cases below, where the first two are necessarily 
singular due to the Weitzenbock formula:

{\bf (i)} $b>1~(n \geq 3)$ : These solutions are homeomorphic to the 
sphere ($R^{1}_{\: 212} \geq 0$) except for a line of singularities along 
the curve $|g| = 1$; we consider them no further.

{\bf (ii)} $b=1~(n =2)$ : The curvature $R^{1}_{\: 212}$ now vanishes, 
but the $U(1)$ curvature $2$--form $F$ does not. We have shown 
\cite{art4} earlier that elliptic or hyperelliptic curves of the form 
$g^{2}(z)~=~(z-a_1)(z-a_2)\ldots(z-a_k)$ lead to solutions with $k$ 
singular vortices centered at the locations 
$a_1 \, , \, a_2 \, , \ldots \, , \, a_k$ on the complex plane \IC. 
Alternatively, one may \cite{art4} choose $g^2 (z) = \sigma (z)$, where 
$\sigma$ is Weierstrass's quasi--doubly periodic function defined by 
$\sigma'/\sigma\equiv\zeta(z)\equiv-\int\wp(z)dz$. As this amounts to 
dividing $\bf C$ by the lattice $\Gamma$ of points 
$\omega = n_1 \omega_1 + n_2 \omega_2$ , one ends up with a genus 
$p = 1$ solution with one singular vortex per cell.

{\bf (iii)} $1>b>0~(n = -1,~0,~1)$ : We now have $b = 1/4~,~1/2$ and 
$3/4$. Pursuing the relation between our solutions and the 
classification of Riemann surfaces according to genus, we see the value 
$b = 1/2$ assigns ${\cal M}_2$ the Poincar\'{e} metric 
\begin{equation}\label{67}
ds^2 = \frac{ dg \, d\overline{g}}{(1-g\overline{g})^2}
\end{equation}
\noindent corrresponding to constant negative curvature 
$R^{1}_{\: 212} = -2a^2$. The $SU(1,1)$ symmetry discussed in section 5 
is now an isometry of the metric (\ref{67}). The Klein form of (\ref{67}), 
obtained by replacing $g(z)$ by $f(z)$ via (\ref{60}), leads to the 
well--known metric 
\begin{equation}\label{68}
ds^2 =\frac{ df \, d\overline{f}}{(Im \, f)^2}
\end{equation}
\noindent on the upper--half--plane $\IC_+$. In the terminology of 
\cite{art9} , $p\geq 2$ Riemann surfaces are obtained by dividing 
$\IC_+$ by $SL(2,\Z)$ into $4p$--gons $D_p$ with geodesic sides 
(arcs of circles of finite or infinite radius, intersecting 
${\rm Im} \, f = 0$ at right angles) ; the sides are of course identified 
according to the standard prescription given in, say, \cite{art13}. 
The identification is accomplished by choosing an $f_p (z)$ which maps 
$\IC_+$ to $D_p$~; this means $f_p (z)$ should be an appropriate 
member of the family of Fuchsian functions \cite{art14} introduced by 
Poincar\'{e}. More precisely, $f_p (z)$ is a Fuchsian function of the 
first kind by which an algebraic function, whose Riemann surface has 
genus $p$ , is uniformized \cite{art15} . Thus with $b = 1/2$ , $c = 0$ , 
$g(z)=(f_p (z)-i)/(f_p (z)+i)$ , a compact, closed Riemann surface of 
genus $p \geq 2$ supporting $p-1$ non--singular `magnetic vortices' and 
a constant Weyl spinor solves the SWME. 

Turning to the $SU(1,1)$ transformation (\ref{54})--(\ref{59}) or their 
$SL(2,\IR)$ versions based on (\ref{61}), we see that neither $U(1)$ 
gauge transformations nor conformal changes in the metric are allowed 
when the choice $b = 1/2$ , $c = 0$ is made; all the fields are now simply 
invariant! This fits in perfectly with the above solution based on 
$f_p (z)$: the constant negative curvature hyperboloid (or, equivalently, 
the upper half--plane $\IC_+$) has been tesellated by $4p$--gons 
$D_p$ which transform into each other under the $SL(2,\Z)$ 
subgroup of $SL(2,\IR)$ ; the invariance of the fields under the same 
transformations ensures that all the $D_p$ are completely equivalent.

For completeness, we summarize below some facts concerning 
$f_p (z)$ . Taking the $4p$ vertices of the polygon $D_p$ at the points 
$z = a_{\nu}$ ($\nu = 1, \, \ldots \, , 4p$) on the real axis, $f_p (z)$ 
is given by (in the notation of \cite{art9})
\begin{equation}\label{69}
f_p (z) = \frac{ u_1 (z)}{u_2 (z)}~~,
\end{equation}
\noindent where $u_1 (z)$ and $u_2 (z)$ are the two linearly independent 
solutions of 
\begin{equation}\label{70}
u''(z)+
\left[
\frac{1}{4}\sum_{\nu=1}^{4p}\frac{1-\alpha_{\nu}^2}{(z-\alpha_{\nu})^2} + 
\frac{1}{2}\sum_{\nu=1}^{4p}\frac{\beta_{\nu}}{z-\alpha_{\nu}}
\right]
u(z) = 0~~.
\end{equation}
\noindent In (\ref{70}), $\pi\alpha_{\nu}$ is the angle at vertex $\nu$ and 
the real constants $\beta_{\nu}$ satisfy
\begin{equation}\label{71}
\sum_{\nu=1}^{4p} \beta_{\nu} = 0 \:\: , \:\: 
\sum_{\nu=1}^{4p}(2\alpha_{\nu}\beta_{\nu}+1-\alpha_{\nu}^2)=0\:\:,\:\:
\sum_{\nu=1}^{4p}[\beta_{\nu}\alpha_{\nu}^2+
\alpha_{\nu}(1-\alpha_{\nu}^2)]=0~~.
\end{equation}
\noindent The chief difficulty in obtaining a power--series solution of 
(\ref{70}) lies in relating the $4p-3$ constants $\beta_{\nu}$ to the 
precise geometry of $D_p$ .

Turning finally to the cases $b= 1/4~~(n=-1)$ and $3/4~~(n=+1)$ , we see 
that (\ref{41}) indicates the curvature $R^{1}_{\: 212}$ can have 
singularities or zeroes depending on the choice of $g(z)$. There is no 
obvious geometrical or topological interpretation for these solutions; we 
consider them no further.

\noindent {\bf 7. Concluding remarks : }

The reader may have wondered what kind of solutions would have followed 
had we made the choice $(\psi_1 = 0, \, \psi_2 \neq 0)$ just before 
equation (\ref{15}). It is not difficult to repeat the subsequent steps 
and see that this only results in $A_{\mu} \rightarrow -A_{\mu}$ , 
$F_{12} \rightarrow -F_{12}$ and $(\psi_1 (g,\overline{g}),0) \rightarrow 
(0, \psi_2 (\overline{g}, g))$ ; everything else remains unchanged. 
Recalling the components $\psi_1$ and $\psi_2$ of a Weyl spinor 
correspond to particle and antiparticle states of the same chirality, we see 
that the new solution is simply the charge conjugate of the original one. 
Note that there are no solutions where {\em some} of the vortices are `up' 
and others are `down'.

The examples presented so far certainly do not exhaust the set of $2d$ 
SWME solutions. First of all, even within our narrow Ansatz, there is 
initially no restriction on $h_{\alpha \beta}$ since the SWME do not 
constrain the form of the metric through, say, Einstein's equation in 
General Relativity. Thus one could try to solve (\ref{18}) for an 
arbitrarily assigned conformal function $\phi$. One could also consider a 
more general Ansatz with $A_3$ and/or $A_4$ non--zero, with or without a  
`warp factor' $\mu (x_1 , x_2 )$ modifying the basis one forms $e^3$ and 
$e^4$ from their simple form in (\ref{3}) to $e^3 = \mu dx^3$ , 
$e^4 = \mu dx^4$ . There are reasons to believe that these more general 
versions of the twice--dimensionally reduced SWME will be found to be 
related to $2d$ integrable systems other than those involving the 
Liouville equation. For example the relation of  $N = 2$ supersymmetry 
gauge theory to integrable systems has been pointed out in \cite{art16}, 
\cite{art17} and \cite{art18}; the last reference especially emphasizes the 
Riemann surface structure and Toda dynamics (the simplest example of 
which is the Liouville equation!) underlying reference \cite{art10}. 
Starting from the integrable systems end of the connection, the
non--linearization procedure in \cite{art19}, \cite{art20} and \cite{art21} 
exhibits a formal similarity to equation (\ref{2}), while the 
dimensionally--reduced form of equation (\ref{1}) may be regarded as the 
counterpart of the Lax pair of the equations for 
$\psi_t$ and $\psi_x$ \cite{art22}.

Finally, we would like to draw attention to a number of mathematical features  
shared by the Seiberg--Witten model, \cite{art10} exhibiting quark confinement 
through monopole condensation and the analysis presented here. In \cite{art10}   
the function 
${\cal F}(a)$, which determines the local part of the $N=2$ SUSY 
effective Lagrangian, has a second derivative ${\cal F''}(a)$ which 
coincides with our $f_{1/2}$ . Of course, formally putting $p = 1/2$ in 
(\ref{69})--(\ref{71}) merely provides a definition of the inverse of the 
elliptic modular function and is not meant to imply that we are dealing with 
a surface of fractional genus. In fact, the analysis in \cite{art10} 
is based on a genus one elliptic curve. Among our solutions, on the 
other hand, the non--singular ones of particular interest due 
to their topologically non--trivial properties are those that are closely 
related to higher genus hyperelliptic cases. These higher genus Riemann 
surfaces also appear in the context of $N=2$ SUSY Yang--Mills theory when 
gauge groups beyond $SU(2)$ are considered while our work makes no 
explicit  reference to the original unbroken non--abelian gauge group. 
There also seems to be a parallelism between the function $f(z)$ in our work 
and the parameter $\tau = \theta / 2\pi + 4\pi i / e^2$  in that both undergo 
$SL(2, \Z)$ projective transformations. The elucidation of the deeper reasons 
behind these mathematical coincidences requires further study.

\noindent {\bf Acknowledgements}

We have benefited greatly from discussions with M. Arýk, T. Dereli, E. 
Ferapontov, S. Finashin, H. G\"{u}mral, R. G\"{u}ven, A. Klyachko and 
Y. Nutku.

\end{document}